\documentclass[%
 aip,
 amsmath,amssymb,
 reprint,%
]{revtex4-1}

\usepackage{graphicx}
\usepackage{dcolumn}
\usepackage{bm}
\usepackage{xcolor}
\usepackage[utf8]{inputenc}
\usepackage[T1]{fontenc}
\usepackage{mathptmx}
\usepackage{etoolbox}
\usepackage[siunitx]{circuitikz}
\usepackage{adjustbox}
\usepackage[font = bf]{subcaption}
\usepackage{caption}
\usepackage{tikz}
\usepackage{fancyhdr}

\usetikzlibrary{decorations.pathmorphing}
\tikzset{%
lray/.style={decorate, decoration={
 snake, amplitude=2pt,pre length=1pt,post length=2pt, segment length=5pt,},
 -Triangle,
 }}
\makeatletter
\def\@email#1#2{%
 \endgroup
 \patchcmd{\titleblock@produce}
  {\frontmatter@RRAPformat}
  {\frontmatter@RRAPformat{\produce@RRAP{*#1\href{mailto:#2}{#2}}}\frontmatter@RRAPformat}
  {}{}
}%

\makeatother
\begin{document}

\preprint{AIP/123-QED}

\title{A 4 - 8 GHz Kinetic Inductance Travelling-Wave Parametric Amplifier Using Four-Wave Mixing with Near Quantum-Limit Noise Performance}
\author{Farzad Faramarzi}
 \email{farzad.faramarzi@jpl.nasa.gov.}
\affiliation{Jet Propulsion Laboratory, California Institute of Technology, Pasadena, CA 91101,USA
}%
\author{Ryan Stephenson}%
\affiliation{ Division of Physics, Mathematics and Astronomy, California Institute of Technology, Pasadena, CA 91125, USA}
\affiliation{Jet Propulsion Laboratory, California Institute of Technology, Pasadena, CA 91101,USA }%

\author{Sasha Sypkens }
\affiliation{ School of Earth and Space Exploration, Arizona State University, Tempe, Arizona 85281, USA}
\affiliation{Jet Propulsion Laboratory, California Institute of Technology, Pasadena, CA 91101,USA }%

\author{Byeong H. Eom}
\affiliation{Jet Propulsion Laboratory, California Institute of Technology, Pasadena, CA 91101,USA }
\author{Henry LeDuc}
\affiliation{Jet Propulsion Laboratory, California Institute of Technology, Pasadena, CA 91101,USA }
\author{Peter Day}%
       \email{peter.k.day@jpl.nasa.gov}
\affiliation{Jet Propulsion Laboratory, California Institute of Technology, Pasadena, CA 91101,USA }

\date{\today}

\begin{abstract}
Kinetic inductance traveling-wave parametric amplifiers (KI-TWPA) have a wide instantaneous bandwidth with near quantum-limited noise performance and a relatively high dynamic range.  Because of this, they are suitable readout devices for cryogenic detectors and superconducting qubits and have a variety of applications in quantum sensing. This work discusses the design, fabrication, and performance of a KI-TWPA based on four-wave mixing in a NbTiN microstrip transmission line.  This device amplifies a signal band from 4 to 8~GHz without contamination from image tones, which are produced in a separate higher frequency band. The 4 - 8~GHz band is commonly used to read out cryogenic detectors, such as microwave kinetic inductance detectors (MKIDs) and Josephson junction-based qubits. We report a measured maximum gain of over 20 dB using four-wave mixing with a 1-dB gain compression point of -58 dBm at 15 dB of gain over that band. The bandwidth and peak gain are tunable by adjusting the pump-tone frequency and power. Using a Y-factor method, we measure an amplifier-added noise of $ 0.5 \leq N_{added} \leq 1.5$ photons from 4.5 - 8 GHz.\footnote{© 2024. All rights reserved}

\end{abstract}

\maketitle

\section{Introduction}

Using a quantum noise-limited parametric amplifier as a first-stage amplifier for readout can improve the sensitivity of multiple cryogenic detector technologies, such as the Microwave Kinetic Inductance Detector (MKID) \cite{zorbist}, and microwave SQUID multiplexer ($\mathrm{\mu}$MUX) readout of Metallic Magnetic Calorimeters \cite{mmc} and Transition-Edge Sensors \cite{mumux}. Traveling wave parametric amplifiers are also an appealing choice for fast and high-fidelity readout \cite{peng_naghiloo_wang_cunningham_ye_obrien_2022,barzanjeh_divincenzo_terhal_2014,didier_kamal_oliver_blais_clerk_2015} of cryogenic qubits. A commonly used frequency band to read out superconducting qubits and detectors is 4 - 8 GHz due to its availability, accessibility, and maturity of readout electronic systems and components at a relatively low cost. Applications of superconducting parametric amplifiers also extend to fundamental physics research for laboratory-based experiments such as dark matter searches, where a quantum-noise-limited gain is of interest \cite{karthik}.

The resonant Josephson junction amplifiers (JPAs) are the most commonly used superconducting parametric amplifiers \cite{Aumentado}. Even though these devices have shown quantum-limited noise performance, they have low fractional bandwidths and very low dynamic range\cite{Esposito}. Recently, by implementing a traveling-wave periodic structure\cite{Aumentado,mutus2013design}, the bandwidths of Josephson traveling-wave parametric amplifiers (JTWPAs) have been increased to a few gigahertz \cite{McklinJTWPA}. However, the low dynamic range of these devices still remains an issue, especially for the readout of resonator arrays \cite{mutus2013design}. In addition, the complexity of circuitry of the nonlinear lumped element transmission line and junction fabrication also puts them at a disadvantage compared to their counterparts, the kinetic inductance-based parametric amplifiers\cite{shibo,malnou,eom_day_leduc_zmuidzinas_2012,ranzani2018,vissers2016,chaudhuri2017,shan2016,Bockstiegel2014}.

Kinetic inductance-based superconducting parametric amplifiers utilize the nonlinear kinetic inductance of superconducting thin films such as NbTiN. Four-wave mixing (4WM) processes can occur in kinetic inductance transmission line structures patterned in coplanar waveguide \cite{eom_day_leduc_zmuidzinas_2012} or microstrip line \cite{shibo} geometries when a strong pump tone is present. A phase-matching condition is necessary to obtain maximum gain over a desired bandwidth. Implementing geometric dispersion allows for control of phase matching. This technique, called dispersion engineering, can also be utilized to mismatch unwanted higher-order nonlinear processes such as pump third harmonic\cite{eom_day_leduc_zmuidzinas_2012}.

Three-wave mixing (3WM) is possible with KI-TWPAs when a constant current is applied to the transmission line. 3WM KI-TWPAs have demonstrated a near quantum-limited noise, wide bandwidth, and high dynamic range\cite{nikita,malnou}. However, since the pump tone corresponds to twice the frequency at the center of the gain curve, it is then folded about its center in the sense that a signal on one side of $f_{pump}/2$ produces an idler tone at the reflected frequency of $f_{pump} - f_{signal}$. That situation is far from ideal for reading out densely frequency-spaced arrays of resonators where the resulting frequency collisions would seriously impact the yield and other cases where there is power at frequencies corresponding to the idler part of the band. The total bandwidth of the published 3WM devices is also somewhat less than 4-8 GHz \cite{nikita,malnou}.

This paper presents a degenerate four-wave mixing KI-TWPA, where the pump tone is between the signal and idler frequencies, $2 f_{pump} = f_{signal} + f_{idler}$. The amplifier tends to produce gain in disjointed frequency ranges that we will refer to as the signal and idler (or image) bands, which makes it possible to separate the idler using ancillary circuitry and use the full signal band without contamination from the image (or idler) frequencies. The devices described here are made of NbTiN as the high kinetic inductance material and patterned in an inverted microstrip geometry\cite{shibo}.

\section{Device Design and Fabrication}

\begin{figure}
     \centering
     \begin{subfigure}[b]{0.75\columnwidth}
     \caption{}
         \centering
         \includegraphics[width=\columnwidth]{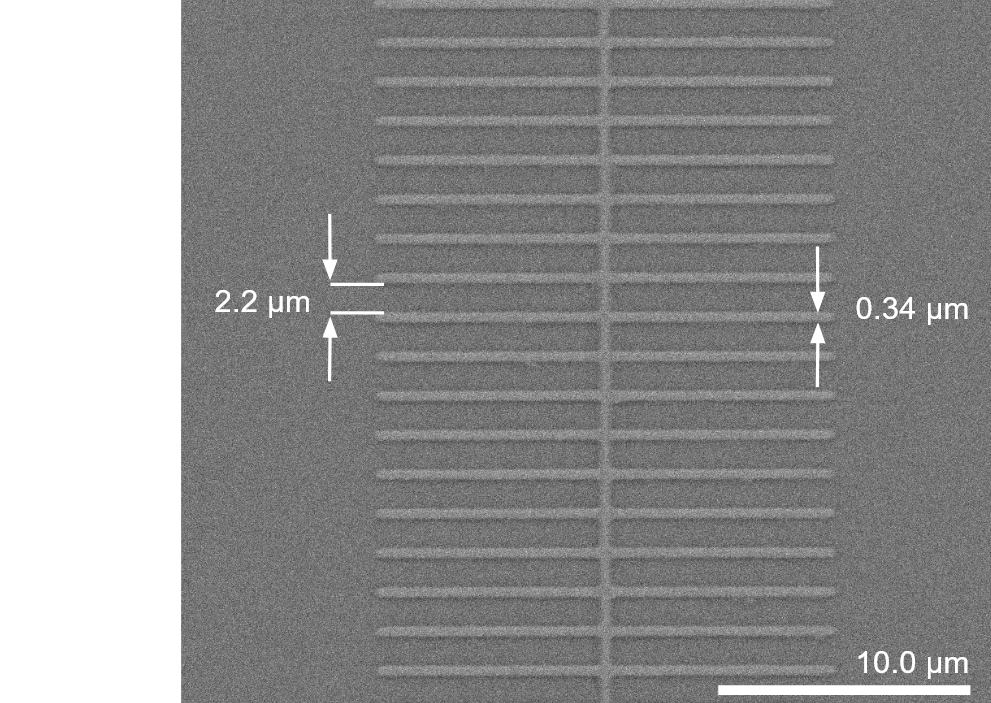}
         \label{fig:a}
     \end{subfigure}
     
     \begin{subfigure}[b]{0.75\columnwidth}
     \caption{}
         \centering
         \includegraphics[width=\columnwidth]{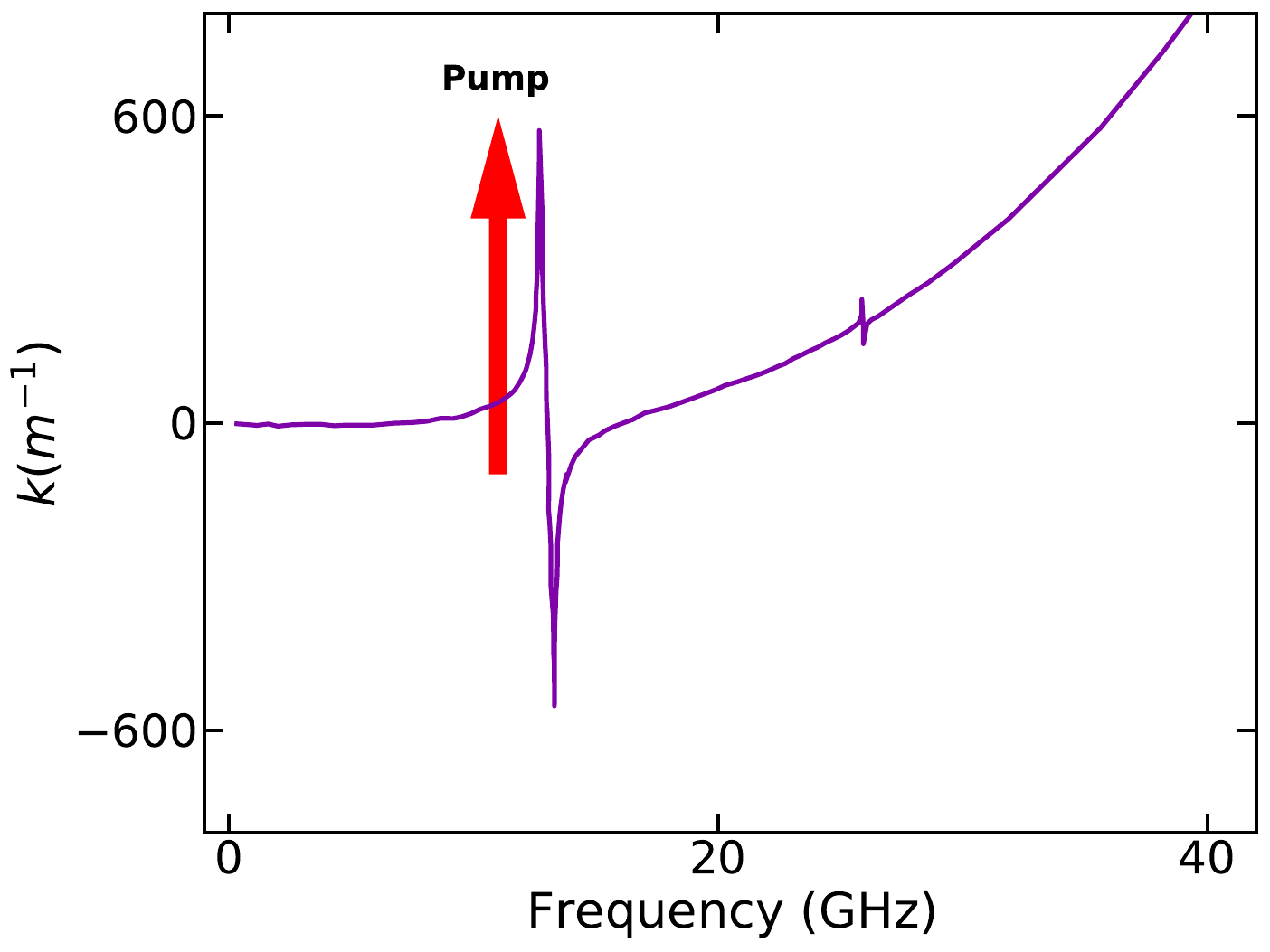}
         \label{fig:b}
     \end{subfigure}
     \begin{subfigure}[b]{0.75\columnwidth}
     \caption{}
         \centering
         \includegraphics[width=\columnwidth]{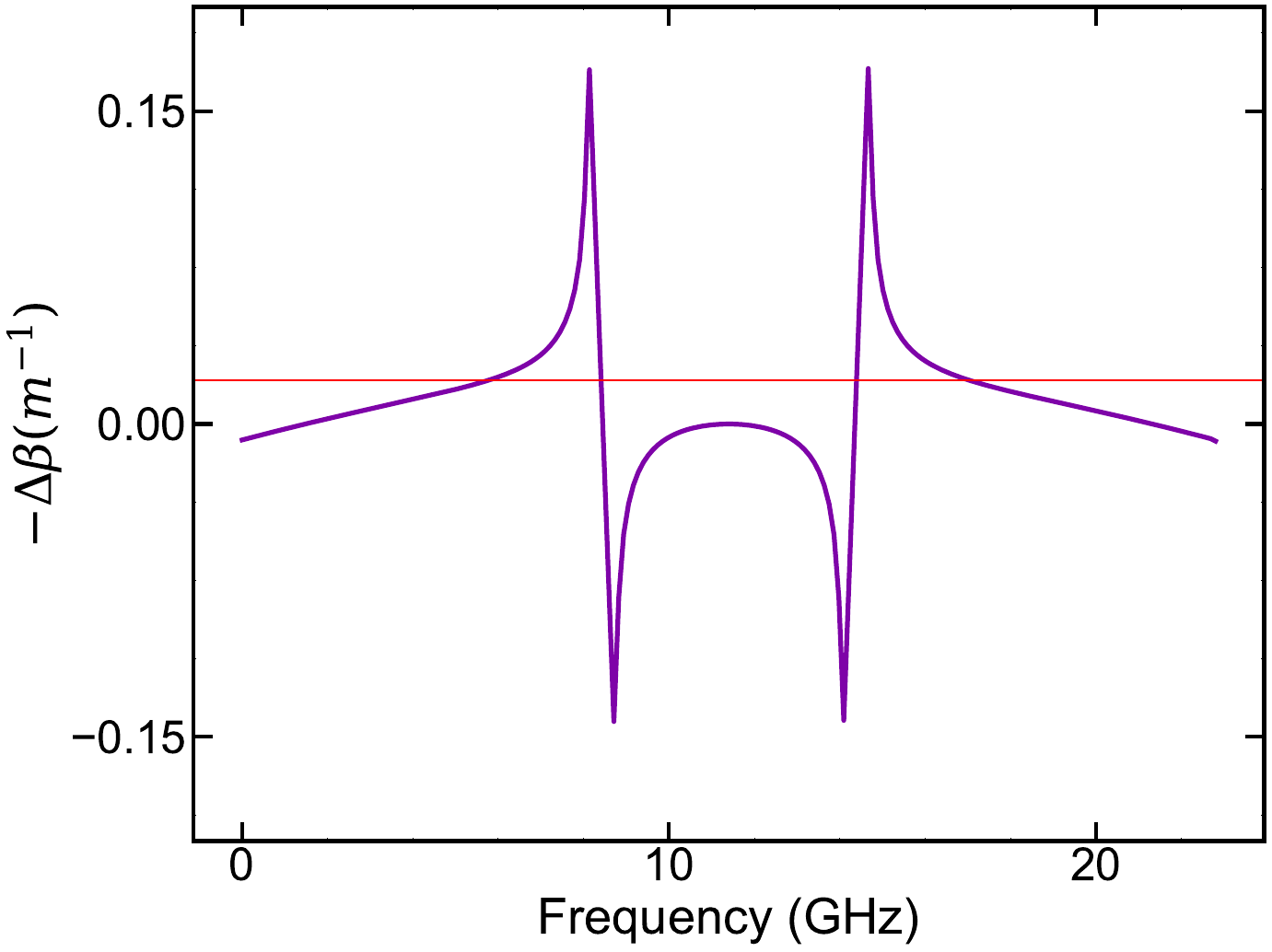}
         \label{fig:b}
     \end{subfigure}
        \caption{\label{disper:gain} a) A SEM micrograph of the transmission line with capacitive stubs. The width of the stubs is 340 nm, the spacing between them is 2.2 microns, and the average length of the stubs is 10.8 microns.  b) Plot of k, the calculated propagation constant, with a linear part subtracted.  The discontinuity at around 12.5 GHz results from the modulation of the length of the capacitive stubs with a periodicity of 122.7 microns. The red arrow shows the position of the pump tone relative to the bandgap. c) Plot of calculated phase matching criterion, $\Delta \beta$, as a function of frequency. The red line is the self-phase modulation of the pump, $\kappa_p I_p^2/(4 I_*^2)$.  The purple line is $\kappa_s + \kappa_i - 2 \kappa_p$. The optimal parametric gain (maximum gain) is achieved near the outer intersecting points.}
        \label{fig:dispersion}
\end{figure}

The design of the device is similar to what was reported in Shu et al. \cite{shibo} with the exception of Nb as the ground plane. To achieve a high nonlinearity in the kinetic inductance of the transmission line as a function of the pump current ($I_p$), the transmission line width was designed to be 340 nm as shown in Fig.~\ref{fig:dispersion}.a. The kinetic inductance per unit length as a function of pump current is given by the following relation\cite{eom_day_leduc_zmuidzinas_2012}:  
\begin{equation}
    \mathcal{L}_k(I_p) = \mathcal{L}_k(I_p=0) \Bigg( 1 + \frac{I_p^2}{I_*^2} + ... \Bigg),
\end{equation}
 where $I_*$ is the ``characteristic'' current that sets the scale of the nonlinearity.

The increase in the impedance of the transmission line, $Z \approx \sqrt{\mathcal{L}_k/\mathcal{C}}$, due to its narrow width and large kinetic inductance was compensated for by adding capacitive stubs to increase the capacitance per unit length,$\mathcal{C}$, of the transmission line. The total length of the device is 100 mm, and the calculated propagation velocity and the characteristic impedance are 0.010c and 50 $\mathrm{\Omega}$, respectively. Using these values, we estimate an inductance per unit length of $\mathcal{L} = 16.64\; \mathrm{\mu H/m}$ and a capacitance per unit length of  $\mathcal{C} = 6.45\; \mathrm{n F/m}$. The length of the stubs was set to create the dispersion needed for the four-wave mixing process and further modulated along the length of the transmission line to create a photonic bandgap \cite{eom_day_leduc_zmuidzinas_2012} around the pump frequency, as shown in Fig~\ref{disper:gain}.b. The average length of the capacitive stubs is 10.8 $\mathrm{\mu m}$ with amplitude modulation of 2.08 $\mathrm{\mu m}$ and periodic modulation of 122 $\mathrm{\mu m}$ to create a bandgap centered at 12.5 GHz.

We can approach the following phase-matching criterion for a wide range of frequencies by putting the pump at the upward step before the bandgap:

\begin{equation}
    \Delta \beta \equiv \kappa_s + \kappa_i - 2 \kappa_p = - \kappa_p \; \frac{I_p^2}{4 I_*^2} \,,
\label{eqn:phasematching}
\end{equation}
where $\kappa_s$,$\kappa_i$, and $\kappa_p$ are the wave vectors corresponding to the signal, idler, and pump tones, respectively, and the term on the right represents the nonlinear dispersive effect \cite{eom_day_leduc_zmuidzinas_2012}.  Fig.~\ref{fig:dispersion}.c shows a plot of $\Delta \beta$ as a function of frequency.

The devices were fabricated by depositing a $35\pm 5$ nm thick NbTiN layer on a 6-inch high resistivity ($\rho \geq 10,000 \; \mathrm{\Omega . cm }$) silicon wafer via reactive sputtering from a NbTi target in a nitrogen atmosphere at ambient temperatures. The NbTiN film was then patterned using a stepper photo-lithography method and etched in a reactive ion etcher. Utilizing a Plasma-Enhanced Chemical Vapor Deposition (PECVD) process, a 100-nm thick amorphous silicon layer was deposited on top of the NbTiN wire. Finally, a 350-nm thick Nb ground-plane layer was sputtered on top of the amorphous silicon layer and then patterned and etched.

\begin{figure}
     \centering
     \begin{subfigure}[b]{0.95\columnwidth}
     \caption{}
         \centering
         \includegraphics[width=\columnwidth]{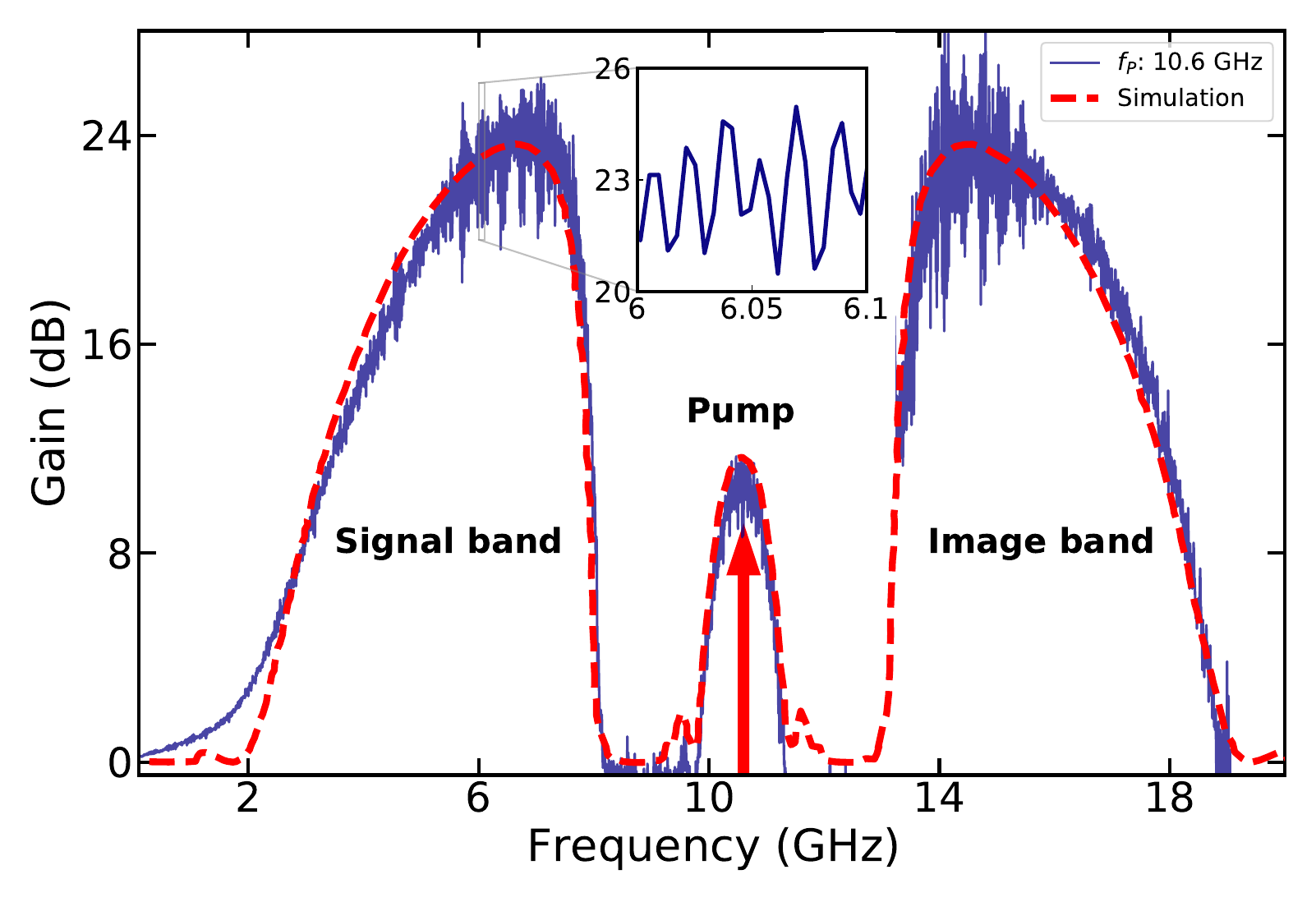}
         \label{fig:a}
     \end{subfigure}
     
     \begin{subfigure}[b]{0.95\columnwidth}
     \caption{}
         \centering
         \includegraphics[width=\columnwidth]{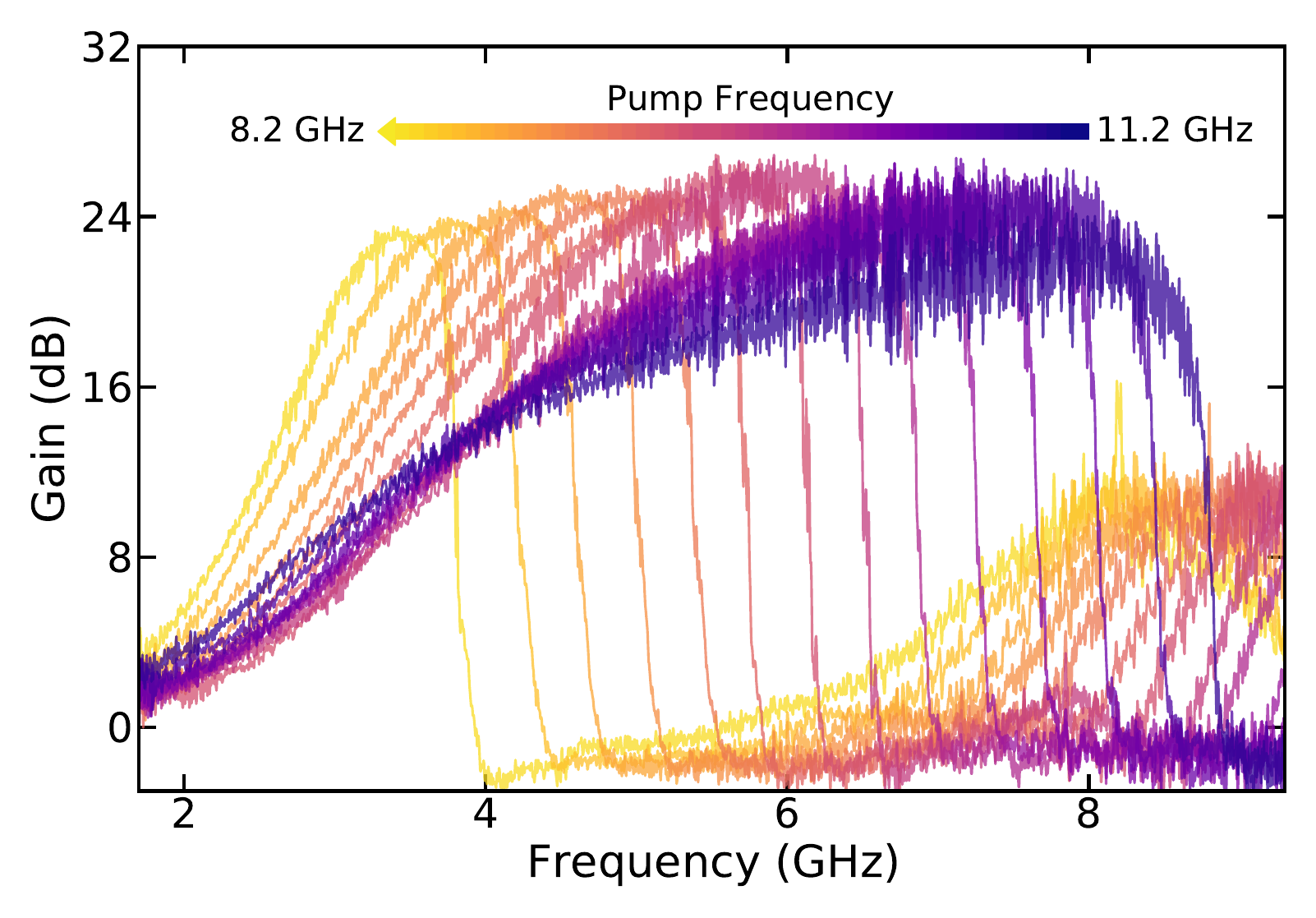}
         \label{fig:b}
     \end{subfigure}
        \caption{\label{fig:gain} a) Measured gain of the device at 1K using a simplified cold measurement setup consisting of only a 20 dB attenuator at 1K before the KI-TWPA and input and output transmission lines directly to room temperature.  The 10.6~GHz pump tone is injected through a directional coupler at room temperature. The frequencies over which the device produces gain are split into three bands.  The gap at around 12.5~GHz is due to the ``engineered'' band gap produced by the periodic modulation of the transmission line structure\cite{shibo,nikita}. The dashed line shows the simulated gain curve and the inset shows the gain ripples in a narrow frequency range. b) Plot of measured gain of the device at 1 K with different pump frequencies. The bandwidth of the ``signal'' band can be adjusted significantly by varying the pump frequency. Measurements shown in this sub-figure are from another device on the same wafer with an identical design.}
        \label{fig:gain}
\end{figure}

\section{Results and Discussion}

\subsection{Behavior}

The gain is measured by normalizing the pumped transmission by the pump-off transmission.  We expect negligible loss from the KI-TWPA based on results reported in Shu et al.\cite{shibo} for a similar device geometry in this frequency range. The KI-TWPA tends to produce gain in disjointed frequency ranges that we refer to as the signal and idler bands (see Fig.~\ref{disper:gain}.a). The amplifier also produces some gain in a narrow frequency band around the pump frequency, but the gain there is less because, in that frequency range, the nonlinear dispersive effect is not compensated by the engineered dispersion. The gap in the gain curve around 12.5~GHz is simply the result of the transmission gap due to the modulation of the stub lengths.  The additional gap around 8.7~GHz occurs because with the signal tone near that frequency, the idler tone would fall within the bandgap, and therefore, the gain process is inoperative.  The gaps in the gain curve serve a useful function in that it is easier to use a diplexer to separate out the pump and the signal and idler bands.  Typically, diplexers have poor input matching at their crossover frequencies, so by placing those frequencies in the zero gain regions, potential problems with reflections can be avoided.   

The expected gain of the device was calculated using the low power dispersion extracted from the measured unpumped transmission.  The measurement was compared to a network model of the KI-TWPA to determine the film surface inductance. In our method, described in ref.~\cite{shibo}, the modulated transmission line is considered to be a uniform medium with an effective dielectric constant, which we derive from the adjusted network model.  That effective dielectric constant is used in the coupled-mode equations to predict the gain.  For the simulated gain curve shown in Fig.~\ref{fig:gain}.a, the higher-frequency products at $2\omega_p+\omega_s$, $3\omega_p$, and $4\omega_p-\omega_s$ were included in the calculation in addition to the signal, pump, and idler frequencies\cite{shibo}. To match the measurement, we set the pump current in the simulation to $I_P = $380 $\mathrm{\mu A}$. The $I_\ast$ value used was 3.2 mA, which was determined by measuring the shift in the stopband as a function of DC current similar to the measurements in Shu et al. ~\cite{shibo}, which results in a ratio of the pump current to the characteristic current of $I_p/I_* = 0.119$. The pump power for the measured gain curve in Fig.~\ref{fig:gain}.a is -20.5 $\pm$ 1.5 dBm, where the error estimate comes from uncertainty in the cryogenic losses in the input line.  The corresponding root-mean-square current is $I_p = 421 \pm 70 \;\mathrm{\mu A}$, close to the value used for the simulation. Some discrepancy between those values may be expected because our model neglects the weakly standing wave nature of the pump that arises from the periodic modulation of the structure. The close agreement between the overall shape and magnitude of the measured and simulated gain curves shown in the figure nevertheless demonstrates that the model captures the main processes occurring in the device and has useful predictive properties.

The frequency band over which the device produces gain can be tuned significantly by adjusting the frequency of the pump tone, as shown in Fig.~\ref{fig:gain}.b.  As the dispersion introduced by the periodic modulation changes rapidly below the bandgap (Fig.~\ref{fig:dispersion}.b), small pump frequency changes result in somewhat larger changes in the signal and idler frequencies that satisfy the phase matching condition, equation (\ref{eqn:phasematching}). 

The measured gain has a rapid frequency variation, shown in Fig.~\ref{fig:gain}.a . The spacing between the gain ripples is 15 MHz, corresponding to a free space length of $\sim$ 20 m, which is twice the electrical length of the KI-TWPA. Therefore, we attribute the ripples to impedance mismatches between the KI-TWPA chip and the feedline circuitry on both sides of the device.  Fig.~\ref{fig:compoint}.b shows the measured gain as a function of pump power with the KI-TWPA inserted in the noise measurement setup of Fig.~\ref{fig:loss}.  Compared to Fig.~\ref{fig:gain}, the ripple amplitude is higher due to the return losses of the additional components.  The increase in the ripple amplitude with pump power can be understood by considering a medium with one-way amplitude gain $g$ between two discontinuities represented by reflection and transmission coefficients $r$ and $t$.  The forward transmission of the structure is 
\begin{equation}
S_{21} = \frac{t^2 g e^{-ikL}}{1 - r^2 g e^{-ikL}},
\label{eqn:S21}
\end{equation}
where $k$ is the propagation constant in the medium and $L$ is the length.  The ripple level increases with $g$ and diverges at $g = r^{-2}$. 

To measure gain compression, a test tone in the middle of the device's signal band was injected into the KI-TWPA, and the system's output power was measured using a spectrum analyzer. With the pump frequency set to 10.6 GHz and pump tone power of -23 dBm, resulting in 15 dB of gain, the output power of the amplified test tone at the spectrum analyzer was measured as a function of the test tone's power at the device.  As shown in Fig.~\ref{fig:compoint}.a, the gain compresses by 1 dB when the input signal is $\simeq$ -58 dBm, corresponding to an output power of -43 dBm. As the power of the output signal and idler tones is provided by the pump, compression occurs when the pump level is significantly depleted.  For our device, that occurs with an output signal power of about 20~dB below the pump power.

\begin{figure}
     \centering
     \begin{subfigure}[b]{0.95\columnwidth}
     \caption{}
         \centering
         \includegraphics[width=\columnwidth]{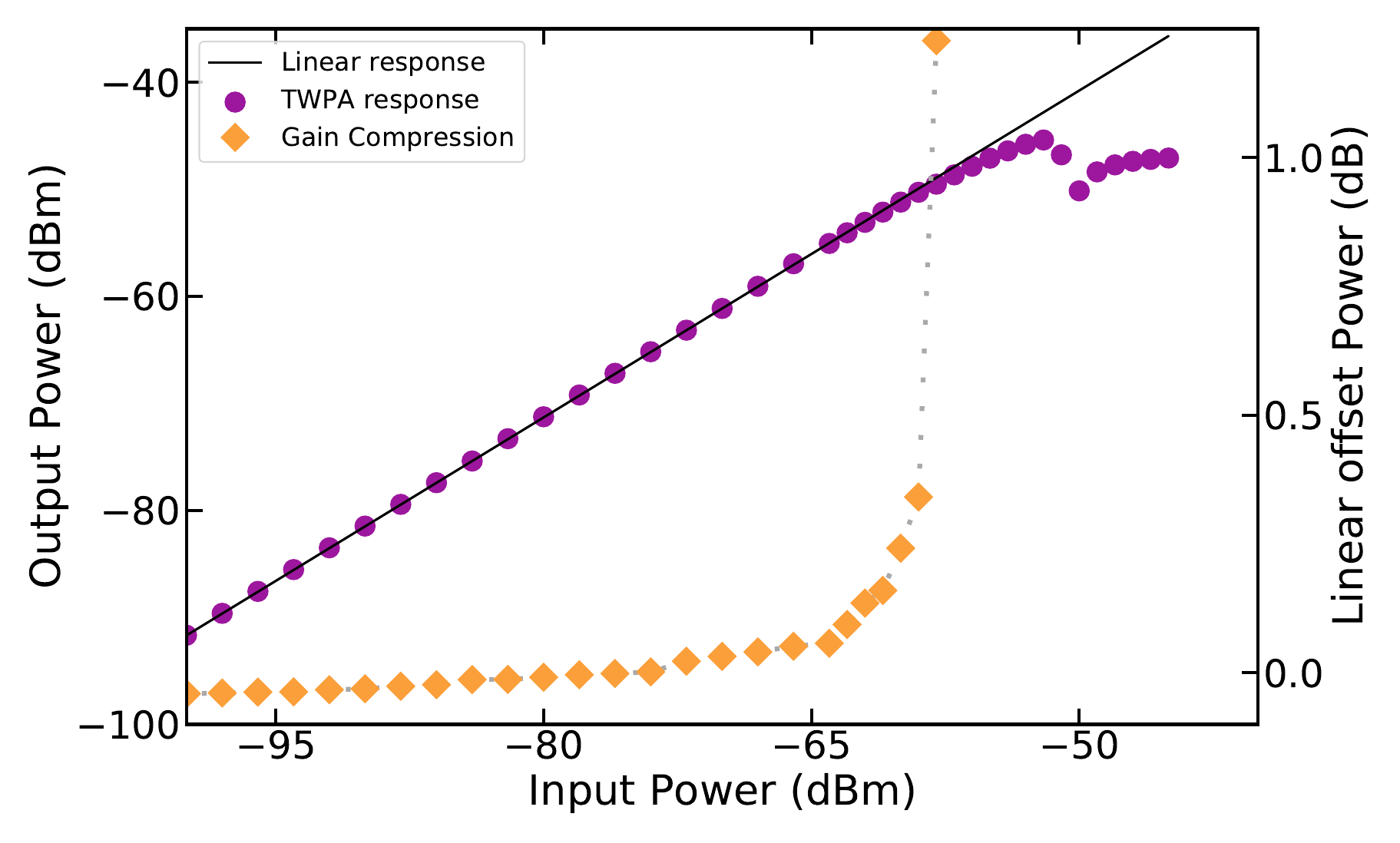}
         \label{fig:a}
     \end{subfigure}
     
     \begin{subfigure}[b]{0.95\columnwidth}
     \caption{}
         \centering
         \includegraphics[width=\columnwidth]{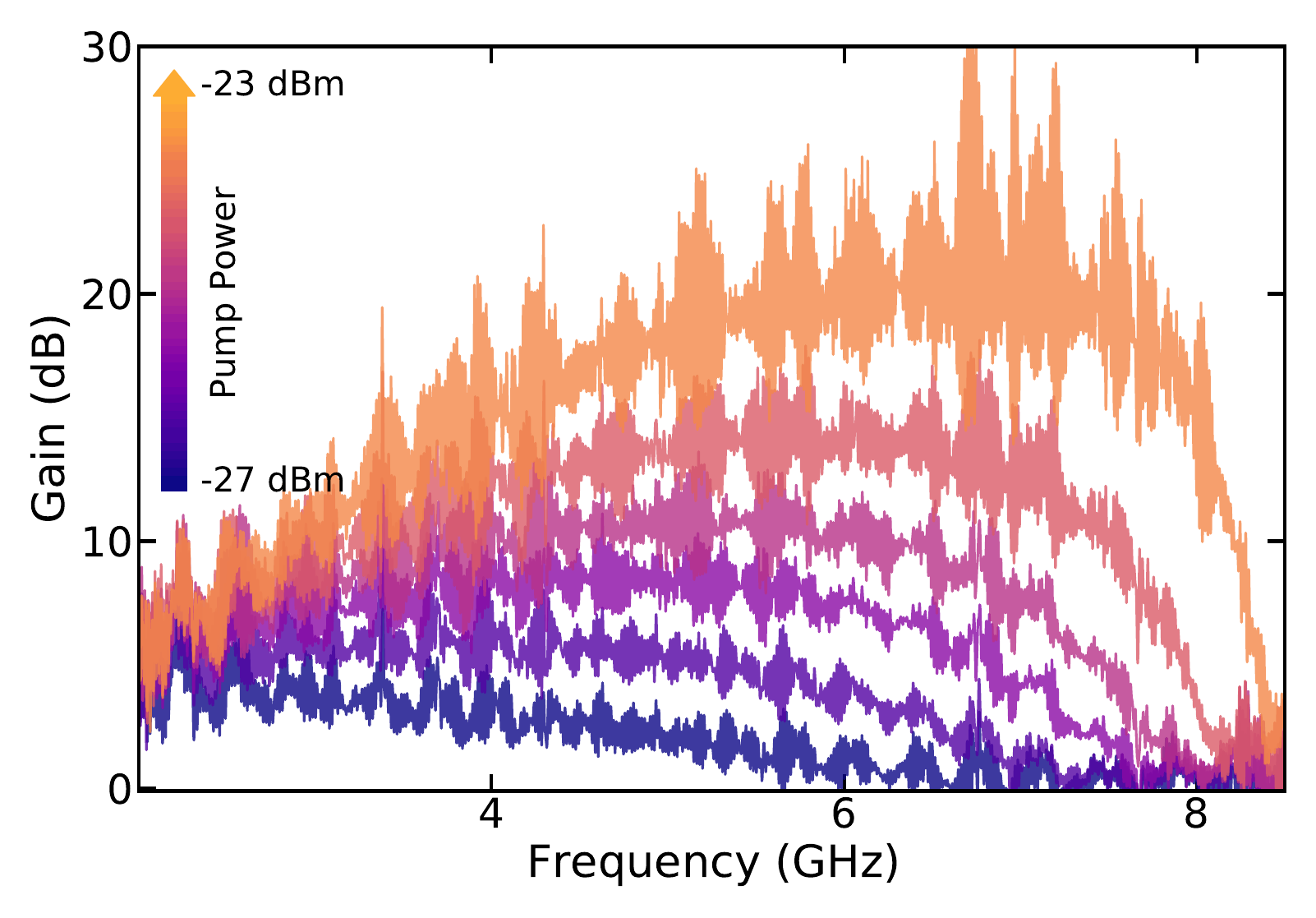}
         \label{fig:b}
     \end{subfigure}
        \caption{\label{fig:compoint} a) Plot of the measured output power at the device of a test tone as a function of the test tone's input power at 20 mK with a signal gain of 15 dB. The gain compresses by 1 dB from the linear response when the input signal power reaches -58 dBm. b) Plot of the measured gain for a fixed pump tone frequency as a function of pump tone power at 20 mK. Both of these measurements were taken using the setup shown in Fig.~\ref{fig:loss}. The higher ripple level in this plot compared to Fig.~\ref{fig:gain}.a is due to the addition of more components in the measurement setup, such as the diplexers and more cabling, which can cause impedance mismatch. }
        \label{fig:compoint}
\end{figure}

\begin{figure*}
     \centering
     \begin{minipage}{0.49\textwidth}
     \begin{subfigure}[b]{0.98\textwidth}
     \caption{}
         \centering
            \begin{adjustbox}{scale=0.48}
            \begin{circuitikz}[american]
            \ctikzset{resistors/scale=0.5,bipoles/length=2cm}
            \draw[color=red]
            (-6.5,5.68) to[R=$T_H$](-4.25,5.68)
            (-6.5,5.68) node[ground,rotate=-90,xscale=0.6]{}
            (-7.5,6.25)rectangle(-4.3,5.3);
            \draw[color=blue]
            (-6.5,4.33) to[R=$T_C$](-4.2,4.33)
            (-6.5,4.33) node[ground,rotate=-90,xscale=0.6]{}
            (-7.5,4.9)rectangle(-4.3,3.95);
            \draw
            (-3,5) node[spdt,xscale=-1.5,yscale=1.5]{}
            (-2.5,5) -- (-1,5)
            (0,7) to [twoport,t = DPX1] (0,9)
            (0.7,8) to[R] (2.5,8)
            (2.5,8) node[ground]{}
            (0,7)--(0,5.67)
            (-1,5) to [twoport,t = DPX2] (1,5)
            (1,5) --(2,5)
        
            (1.3,5.2)node[]{C.O.M}
            (-1.4,5.2)node[]{$<$ 9 GHz}
            (-0.7,6)node[]{$>$ 9 GHz}
            (0.6,7)node[]{C.O.M}
            (7.5,5.2)node[]{$<$ 9 GHz}
            (5.2,6)node[]{$>$ 9 GHz}
            (4.65,5.2)node[]{C.O.M}
            (-1.5,9.5)node[color=violet]{Pump in}
            (4.4,9.5)node[color=violet]{Pump out}
            (0.4,9)node[]{LP}
            (1,8.3)node[]{HP}
            (8.5,4.5)node[]{to isolator}
            (-2.6,5.7)node[]{S}
            
            (2,5) to [amp,t = TWPA] (4,5)
            (4,5)--(5,5)
            (5,5) to [twoport,t = DPX3] (7,5)
            (6,5.7)--(6,7)
            ;
            \draw [dashed] (7,5) to (8.5,5);
            \draw [dashed] (0,9) to (0,10);
            \draw[dashed](6,7) to (6,10);
            \draw [lray,color=violet] (-0.5,10) to (-0.5,9);
            \draw[lray,color=violet] (5.5,9) to (5.5,10);
            
            \end{circuitikz}
            \end{adjustbox}
         \label{fig:a}
     \end{subfigure}
     \\
     \begin{subfigure}[b]{0.98\textwidth}
     \caption{}
         \centering

         \begin{adjustbox}{scale=0.48}
         \begin{circuitikz}[american]
            \draw
            (-2,0) node[ground]{}
            (-2,0) to [vsourceN,l=Noise Source] (0,0)
            (0,0)--(0.5,0)
            (0.5,0) to [generic,l_=$L_1$] (2.5,0)
            (2.5,0) --(3,0)
            (3,0) to [amp,t={PA},l_=$G_{PA}\; N_{A}$] (5,0)
            (5,0)--(5.5,0)
            (5.5,0) to [generic,l_=$L_2$] (7.5,0)
            (7.5,0) -- (8,0)
            (8,0) to [amp,l_=$G_{HEMT}\; N_{HEMT}$](10,0)
            (10,0)--(10.5,0)
            (10.5,0) to [amp,l^=$G_w$] (12.5,0)
            (12.5,0)--(13,0)
            (13,0) to[twoport,t={SA}] (14.1,0)
            ;
            \draw[black,thick,dashed] (0.5,1.5)--(0.5,-1.5)
            (0.125,1.5)node[]{$N_{in}$};
            \draw[black,thick,dashed] (3,1.5)--(3,-1.5)
            (2.7,1.5)node[]{$N_1$};
            \draw[black,thick,dashed] (5.5,1.5)--(5.5,-1.5)
            (5.125,1.5)node[]{$N_2$};
            \draw[black,thick,dashed] (7.5,1.5)--(7.5,-1.5)
            (7.125,1.5)node[]{$N_3$};
            \draw[black,thick,dashed] (10.5,1.5)--(10.5,-1.5)
            (10.125,1.5)node[]{$N_4$};
            \draw[black,thick,dashed] (12.75,1.5)--(12.75,-1.5)
            (12.25,1.5)node[]{$N_{tot}$};
                    
            \end{circuitikz}
            \end{adjustbox}
        \label{fig:b}
     \end{subfigure}
     \end{minipage}
     \hfill
     \begin{minipage}{0.49\textwidth}
     \begin{subfigure}[b]{0.98\textwidth}
     \caption{}
         \centering
         \includegraphics[width=\textwidth]{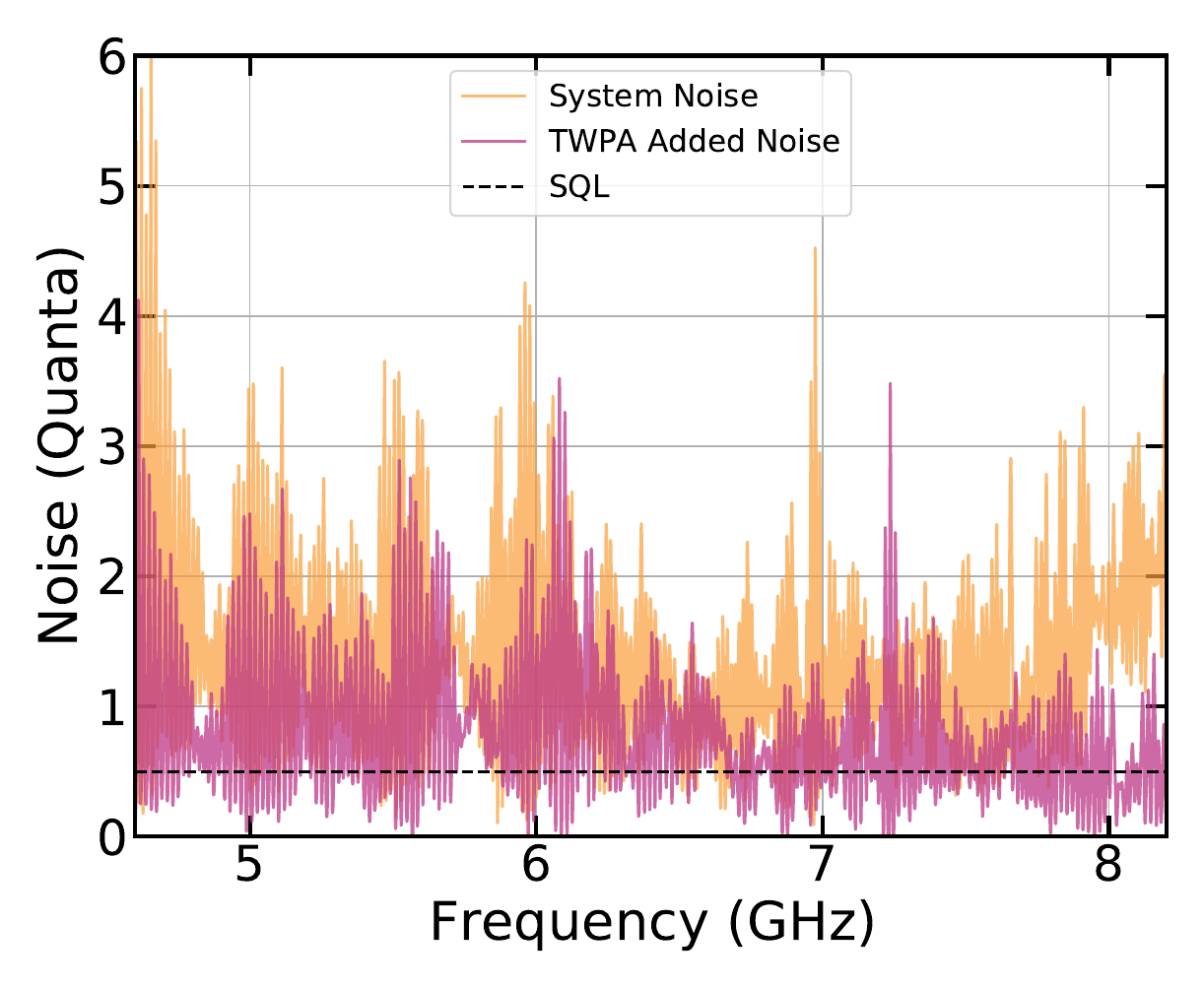}
         \label{fig:b}
     \end{subfigure}
     \end{minipage}
        \caption{\label{fig:noise} a) Schematic of the noise measurement setup with hot and cold broadband noise sources. Two 50 $\mathrm{\Omega}$ cryogenic loads were used as noise sources. The hot source was heat sunk to the 3 K stage of the dilution fridge with a base temperature of 3.18 K, and the cold noise source was placed on the fridge's mixing chamber (MXC) with a base temperature of $\simeq 20 $ mK. All other components in this figure were placed in the MXC. Using a relay switch S, the input of the KI-TWPA was switched between the hot and cold sources. The two types of diplexers have crossovers in the lower and upper gaps in the gain curve shown in Fig.~\ref{fig:gain}.a and are used to inject and separate out the pump and idler tones. b) circuit diagram of the cascaded noise model for the KI-TWPA measurement setup. c) Plot of the system noise and KI-TWPA added noise in units of quanta in the 4.6 - 8 GHz range. The standard quantum limit (SQL) is shown with a dashed line. }
        \label{fig:noise}
\end{figure*}
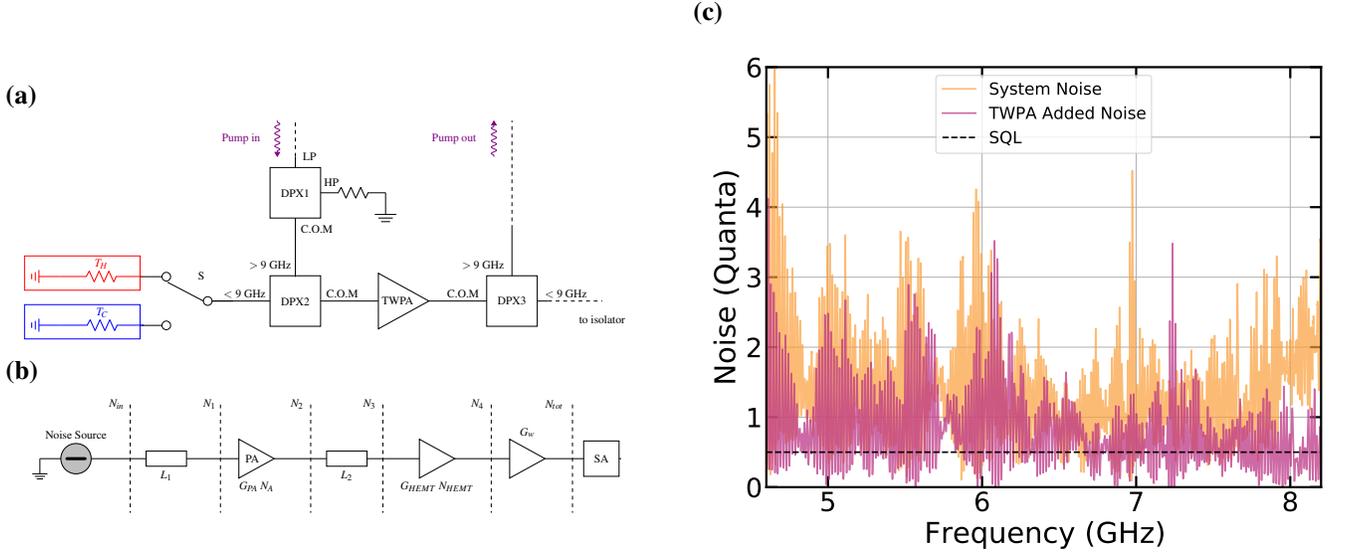

\subsection{Noise}
The amplifier chain noise was measured using a Y-factor method with two 50 $\mathrm{\Omega}$ cryogenic loads as noise sources, one at 3.18 K and the other at the Mixing Chamber (MXC) stage ($\sim$ 20 mK) of the dilution refrigerator. Using a cryogenic relay switch, we alternate between the noise sources as shown in Fig.~\ref{fig:noise}.a. Due to a lack of output-to-input isolation, reflections must be avoided at either side of the TWPA at frequencies where the amplifier has gain. If the sum of the dB return losses on the input and output sides does not exceed the gain, an oscillation will develop. The circuit shown in Fig.~\ref{fig:noise} provides an impedance match at both the input and output of the KI-TWPA over the signal and idler frequency ranges using a combination of diplexer (DPX) circuits. The input side diplexers allow for the injection of the pump tone through a bandpass filter (not shown) that removes room temperature thermal noise and synthesizer phase noise in signal and idler bands that the diplexers may not sufficiently attenuate.  The input side DPX 1 provides a cold termination to the input of the KI-TWPA over the idler band ($>$13~GHz).  The noise from that termination determines the added noise of the KI-TWPA, so it is kept at a temperature $T \sim 20\,$mK $\ll \hbar \omega_{i} / k_B$ for the system to be quantum limited.  In this configuration, the KI-TWPA only receives the thermal noise emitted from the calibration sources in the signal band.

On the output side of the KI-TWPA, the diplexer removes the pump and idler tones and terminates them.  To minimize heating, the pump is terminated at a higher temperature stage ($\sim$ 50 mK).  The idler is also terminated at 50 mK to keep the KI-TWPA quantum limited.  Finally, an isolator is used to avoid noise from the HEMT, which may otherwise contaminate the KI-TWPA via the same mechanism. The signal is then further amplified using a room temperature amplifier and measured using a spectrum analyzer. 

The total noise of the amplifier chain, which includes the KI-TWPA, a Low Noise Factory cryogenic HEMT amplifier (Model: LNF-LNC0.3\_14B), and a room temperature low noise amplifier, was determined using the Y-factor method,
\begin{equation}
    N_{sys} (\omega) = \frac{N_H (\omega)- Y \;N_C(\omega)}{Y-1} ,
\end{equation}
where $\omega$ is the signal angular frequency, $k_B$ is the Boltzmann constant, and $\hbar$ is the reduced Planck's constant. The $N_C(\omega)$ and $N_H(\omega)$ are the noise temperatures of the hot and cold terminations in units of quanta given by \cite{kerr}

\begin{equation}
    N_{H,C} = \frac{1}{2 } \coth \Big( \frac{\hbar \omega}{2k_B T'_{H,C}}\Big)
\end{equation}
where $T'_{H,C}$ is the physical temperature of the noise sources.

The noise added by the KI-TWPA itself was determined using the measured gain of the device and the total system noise from the Y-factor measurement. Microwave losses in the measurement setup, especially between the KI-TWPA and the noise sources, can cause an overestimation of the measured noise. We estimated the loss from the diplexers by comparing the measured transmission through the KI-TWPA and diplexers to that of a bypass cable as shown in Fig.~\ref{fig:loss}.a. After considering the loss, the added noise, $N_A$, in units of quanta, of the KI-TWPA can be calculated as

 \begin{equation}
   \begin{aligned}
    N_A = L_1 N_{sys} - N_{qm}\; &\Bigg( \frac{L_2 \;G_{PA}(1-L_1) + (1-L_2)}{L_2 \; G_{PA}} \Bigg) \\
    &-\frac{N_{HEMT}}{L_2 G_{PA}}
    \end{aligned}
    \label{eqn:addednoise}
 \end{equation}
where $L_1$ and $L_2$ are the losses in the components before and after the KI-TWPA, respectively. $N_{qm}= 0.5 $ is the quantum noise, $N_{HEMT}$ is the added noise of the HEMT amplifier and $G_{PA}$ is the measured gain of the KI-TWPA. The noise model used to arrive at equation (\ref{eqn:addednoise}) is explained in Appendix B.  We measure a total system noise of  $ 1 \le N_{sys} \leq 3$ in the 4.6 - 8 GHz range as shown in Fig.~\ref{fig:noise}.c. Using equation~(\ref{eqn:addednoise}), we estimate the added noise of the KI-TWPA to be $ 0.5 \le N_{A} \leq 1.5$  as shown in Fig.~\ref{fig:noise}. The cryogenic HEMT amplifier's noise was measured using the Y-factor with the pump off. We measured an average noise temperature of 5 K or $\sim$ 13 quanta in the 4 - 8 GHz band consistent with data provided in the amplifier's datasheet. It was observed that the measured added noise in the KI-TWPA deviates from half a quanta in some places (Fig.~\ref{fig:noise}.c). These discrepancies could be attributed to some nonidealities in the test setup. For instance, both noise sources are connected to the input of the KI-TWPA through three cables and two switches. Estimating the insertion loss in all these components is quite challenging, and any unaccounted losses can lead to uncertainty in the measurement and overestimation of the measured noise. Impedance mismatch between the noise terminations and the KI-TWPA is also another source of uncertainty in our noise measurements. The ripples in the measured system noise and the TWPA-added noise are indications of these imperfections. 


\section{Conclusion}
In conclusion, we have designed, fabricated, and measured a wide-bandwidth Four-Wave Mixing KI-TWPA that produces gain between 3 - 9 GHz with a 1 dB compression of -58 dBm suitable for the readout of large detector arrays and superconducting qubits. Using a Y-factor method, we have demonstrated a near quantum-limited noise performance of KI-TWPA in the 4.6 to 8 GHz frequency range.
The amplifier produces gain in disjointed frequency ranges that we call the signal band and a higher-frequency idler (or image) band. This makes it straightforward to separate the idler tones using ancillary circuitry and use the full signal band without contamination from the image (or idler) frequencies for the readout of cryogenic detectors. 
The signal band peak gain of this device is shown to be tunable with the frequency of the pump tone. This tunability makes the KI-TWPA suitable for applications where the signal frequency range needs to be adjusted, such as in dark matter experiments. In addition, the higher frequency idler band of the KI-TWPA can be used to look for dark matter candidates with a higher mass range. The simplest way to accomplish dual-band operation without changing the isolator and second-stage HEMT amplifier would be to operate the KI-TWPA in a frequency translating mode, relying on the fact that the idler band contains a redundant copy of the signal band and visa versa.  A dual-band circuit could be implemented by adding two switches, one before the HP port of DPX1 and the other before the LP port of DPX2 to switch between either signal inputs or terminations.
This capability makes it possible to cover a wide range of masses with a single KI-TWPA.

\begin{acknowledgments}
This research was carried out at the Jet Propulsion Laboratory under a contract with the National
Aeronautics and Space Administration (80NM0018D0004). F.F’s research was supported by appointment to the NASA Postdoctoral Program at the Jet Propulsion Laboratory, administered by Oak Ridge Associated Universities under contract with NASA.
\end{acknowledgments}

\appendix

\begin{figure*}
     \centering
     \begin{subfigure}{0.48\textwidth}
     \centering
     \caption{}
     \includegraphics[width=\textwidth, trim={0 0.2cm 0 0.7cm}, clip]{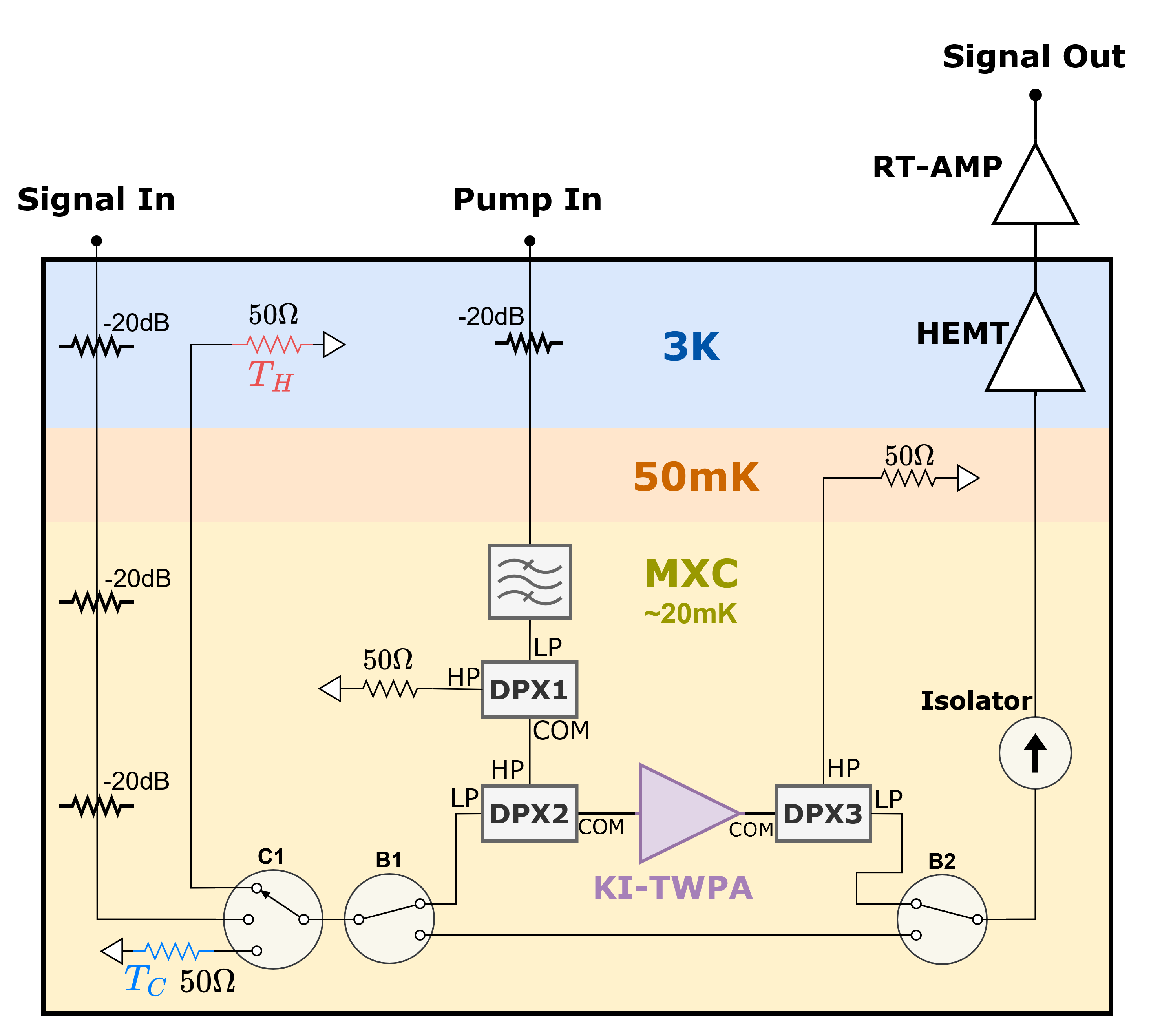}
         \label{fig:a}
 \end{subfigure}
     \hfill
     \begin{subfigure}{0.5\textwidth}
     \caption{}
         \centering
         \includegraphics[width=\textwidth]{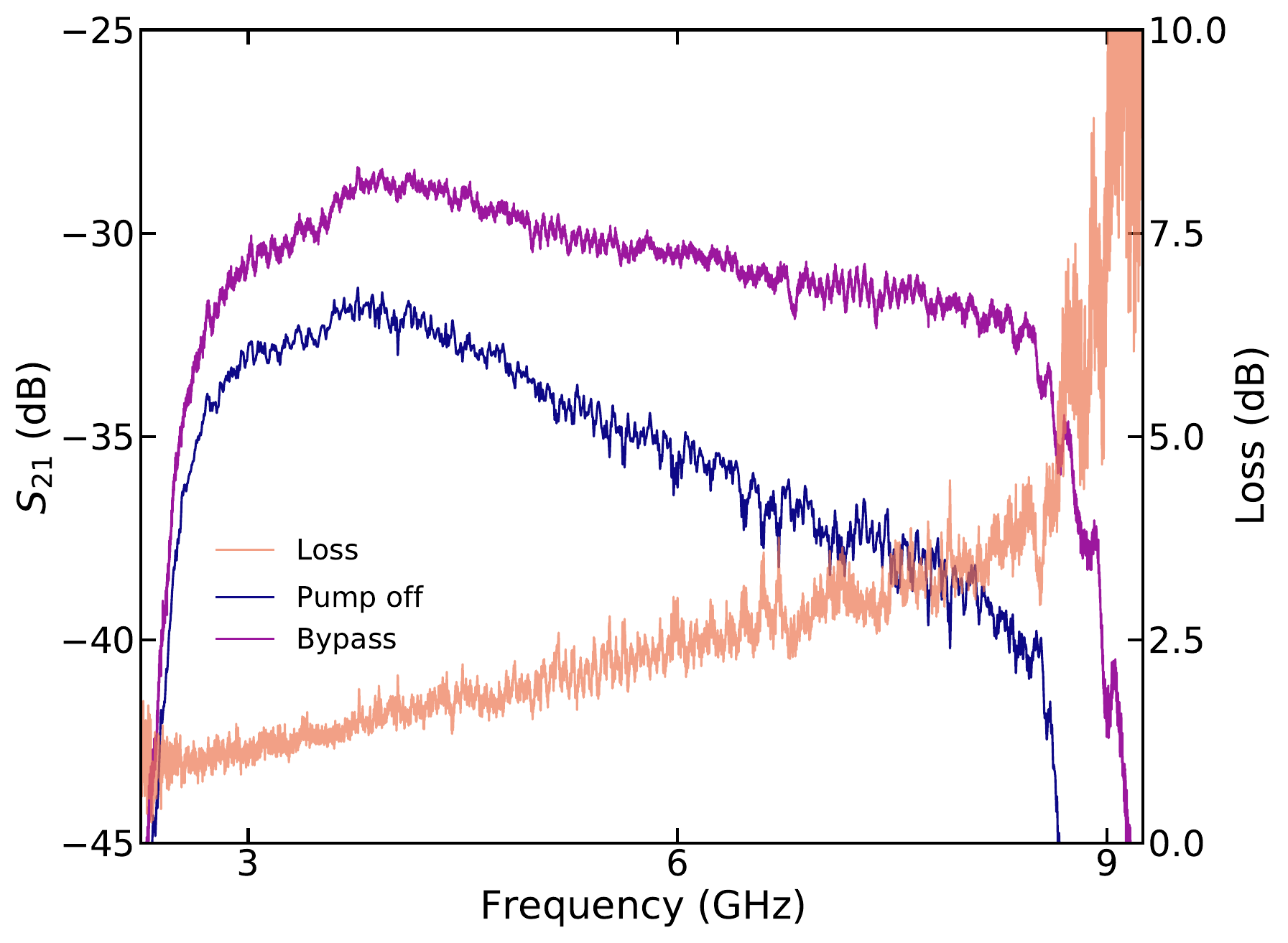}
         \label{fig:b}
     \end{subfigure}
        \caption{\label{fig:loss}a)  Circuit diagram of the noise measurement setup. The input line is attenuated using three cryogenic attenuators and then connected to switch C1. Two cryogenic 50 $\mathrm{\Omega}$ loads were used as noise sources. The `Hot' noise source was placed at the 3 K stage of the dilution fridge, and the `Cold' noise source was placed at the MXC. In this configuration, we can switch between the 50 $\mathrm{\Omega}$ for Y-factor measurements and the signal line for gain measurements. The Pump tone was injected in the KI-TWPA through a bandpass filter and DPX 1 and 2 with cross-over frequencies of 14 and 9 GHz, respectively. The pump tone is then routed out using DPX 3 (cross-over frequency of 9 GHz) and terminated on the 50 mK stage of the cryostat to prevent the HEMT from saturating. The cables used in the measurement setup connecting switch C1 to the `hot' and `cold' terminations on the 3 K and the MXC stages were NbTi superconducting cables. To calibrate the loss in the components, we used a cryogenic cable between switches B1 and B2 to bypass the TWPA and the diplexers. The output is then read out through an isolator, a cryogenic HEMT amplifier, and a room-temperature amplifier using a spectrum analyzer. b) The plot of measured transmission through the entire noise setup with the KI-TWPA unpumped is denoted as `Pump off,' and through the cryogenic cable bypassing the components is denoted as `Bypass.' The plot labeled `Loss' is the difference between the two measurements divided by two to account for the loss in one set of diplexer and cable since identical components were used on both sides of the KI-TWPA.}
        \label{fig:loss}
\end{figure*}
\section{Noise Measurement Setup}

The input line is sufficiently attenuated from room temperature to the MXC stage using a combination of attenuators placed on the 3 K stage and MXC of the dilution refrigerator. For calibration and y-factor measurements, we utilized three relay switches at the MXC. Switch $C_1$ (Fig.~\ref{fig:loss}) is used to switch between the input of the cryostat and the hot and cold terminations. Switches $B_1$ and $B_2$ are synced up and operated to switch between the KI-TWPA and a cryogenic cable for calibration. The amplified signal then passes through an isolator (on MXC) and a cryogenic HEMT amplifier at the 3 K stage, further amplified by a room temperature low noise amplifier, and finally read out using a spectrum analyzer. 

The pump input was filtered using a bandpass filter with a center frequency of 11.45 GHz and a bandwidth of 300 MHz at the MXC stage and directed to the KI-TWPA using DPX1 (Marki, DPX1114) and DPX2 (Minicircuits, ZDSS-7G10G-S+) as illustrated in Fig~\ref{fig:loss}a. Using the highpass port of DPX3 (Minicircuits, ZDSS-7G10G-S+), the pump tone and the idler tones are terminated at the 50 mK stage of the dilution fridge. The total transmission of the system through the device was measured using a VNA with the pump tone off and the transmission of the cable bypassing the KI-TWPA and the diplexers as shown in Fig.~\ref{fig:loss}.a. The plots comparing the transmission between the cryogenic cable and the unpumped KI-TWPA are shown in Fig.~\ref{fig:loss}.b and labeled as `Bypass' and `Pump off,' respectively. The net loss from DPX2 and cables placed between the KI-TWPA and switch C1 in the circuit is also plotted in Fig.~\ref{fig:loss}.b as `Loss.'

\section{Noise Theory}
To model the noise in our setup, we used the circuit diagram shown in Fig.~\ref{fig:noise}.b. If we assume the mK losses in the components between the noise sources (hot \& cold in this case) and the KI-TWPA are $L_1$ and $L_2$, the cascaded noise equations can be written as follows: 

\begin{align}
    N_1& = L_1 N_{in}(T) + N_{qm} (1- L_1), \\
    N_2& = G_{PA} \Big( N_1 + N_A \Big), \\
    N_3& = L_2 N_2 + N_{qm} (1 - L_2), \\
    N_4& = G_{HEMT} \Big( N_3 + N_{HEMT} \Big), \\
    N_{tot}& = G_w \Big( N_4 + N_w \Big)  \ .
\end{align}

Where $N_{in} (T) = (1/2) \coth \left(\hbar \omega/2k_B T \right)$ is the input noise from the 50\,$\mathrm{\Omega}$ terminations. $N_{qm}$ is the quantum noise and for $k_B T  \ll \hbar \omega$ is equal to one half. $G_{PA}, N_{A}$ and $ G_{HEMT}, N_{HEMT}$ are the gain and added noise of the KI-TWPA and the HEMT amplifier, respectively. $G_w$ and $N_w$ are the additional effective warm amplification and warm amplifier noise, respectively. $N_{tot}$ is the total noise measured at the spectrum analyzer (SA). \\

When the KI-TWPA is pumped, the total noise measured by the SA can be calculated using equations (B1-B5). The overall noise contribution of the warm amplifier was ignored because of the high gain ($\sim$ 38 dB) of the HEMT amplifier and the KI-TWPA ($\sim$ 20 dB).

\begin{equation}
    N_{tot}^{on} = G_{eff} \Big[ N_{in}(T) +N_{PA}^{eff} + N_{HEMT}^{eff} \Big] 
\end{equation}
where $G_{eff}$, $N_{PA}^{eff}$ and $N_{HEMT}^{eff}$ are defined as follows.

\begin{equation}
    G_{eff} = G_w G_{HEMT} L_2 G_{PA} L_1 
\end{equation}

\begin{equation}
    N_{PA}^{eff} = \frac{N_{qm} (1-L_1)}{L_1} + \frac{N_A}{L_1}
\end{equation}

\begin{equation}
    N_{HEMT}^{eff} = \frac{N_{qm} (1-L_2)}{L_2 G_{PA} L_1} +\frac{N_{HEMT}}{L_2 G_{PA} L_1}
\end{equation}
By switching to a hot and a cold load, therefore varying the input noise, we can extract the system noise 
\begin{equation}
    N_{sys} = N_{PA}^{eff} + N_{HEMT}^{eff}
\end{equation}

When the KI-TWPA is unpumped, we can treat it as a lossless transmission line. Using the cascaded noise equations, we arrive at the following equation for the total noise measured at the SA with the pump off.
\begin{equation}
    N_{tot}^{off} = \frac{G_{eff}}{G_{PA}} \Big[ N_{in}(T) +\frac{N_{qm} (1-L_1L_2) + N_{HEMT}}{L_1L_2} \Big]
\end{equation}
If we bypass the KI-TWPA and all the components between the noise source and the HEMT amplifier, the total noise measured at SA is 
\begin{equation}
    N_{tot}^{BP} = G_w G_{HEMT} \Big[  N_{in}(T) + N_{HEMT}   \Big].
\end{equation}

Then, the added noise of the KI-TWPA is 

\begin{equation}
    N_A = L_1 N_{sys} - N_{qm}\; \Bigg( \frac{L_2 \;G_{PA}(1-L_1) + (1-L_2)}{L_2 \; G_{PA}} \Bigg) -\frac{N_{HEMT}}{L_2 G_{PA}}
\end{equation}

In the above equations, the loss factors $L_1$ and $L_2$ can be estimated either by measuring insertion losses of the components separately or using the bypass and pump-off transmission measurements using the relay switches. $N_{HEMT}$ can be estimated using either equation (B12) or (B11) with a y-factor measurement. Finally, measuring the gain of the KI-TWPA, we can calculate the added noise of the TWPA from the above equation. Using the unpumped KI-TWPA noise spectra, we measure an average HEMT noise temperature of 5 kelvin, which agrees with LNF-LNC0.3\_14B datasheet.


In Fig.~\ref{fig:loss}.b, we present the transmission ($S_{21}$) measurements of the cable in two scenarios. The first one is the bypassing of KI-TWPA, represented by the purple curve, and the second one is the transmission through DPX2, KI-TWPA, and DPX3, represented by the dark blue curve. We have calculated the loss between switch B1 and the KI-TWPA by subtracting the unpumped measurement from the bypass measurement and dividing it by two since we used identical diplexers with similar insertion loss. This calculated loss is plotted as an orange curve in Fig.~\ref{fig:loss}.b as a function of frequency, and we have used it as L1 and L2 loss factors in the main text.
 
\bibliography{aipsamp}

\end{document}